**Title:** Inverse design and flexible parameterization of meta-optics using algorithmic differentiation

**Authors:**

Shane Colburn[1,*] & Arka Majumdar[1,2,*]

**Affiliations:**

[1]Department of Electrical and Computer Engineering, University of Washington, Seattle, Washington 98195, USA.

[2]Department of Physics, University of Washington, Seattle, Washington 98195, USA.

*Correspondence to: scolbur2@uw.edu, arka@uw.edu

**Abstract: (150 words)**

Ultrathin meta-optics offer unmatched, multifunctional control of light. Next-generation optical technologies, however, demand unprecedented performance. This will likely require design algorithms surpassing the capability of human intuition. For the adjoint method, this requires explicitly deriving gradients, which is sometimes challenging for certain photonics problems. Existing techniques also comprise a patchwork of application-specific algorithms, each focused in scope and scatterer type. Here, we leverage algorithmic differentiation as used in artificial neural networks, treating photonic design parameters as trainable weights, optical sources as inputs, and encapsulating device performance in the loss function. By solving a complex, degenerate eigenproblem and formulating rigorous coupled-wave analysis as a computational graph, we support both arbitrary, parameterized scatterers and topology optimization. With iteration times below the cost of two forward simulations typical of adjoint methods, we generate multilayer, multifunctional, and aperiodic meta-optics. As an open-source platform adaptable to other algorithms and problems, we enable fast and flexible meta-optical design.

**Introduction:**

Metagratings and metasurfaces[1–5] have generated significant interest in recent years, enabling flat optics[6–22] for manipulating the phase, amplitude, and polarization of incident light. These devices comprise arrays of subwavelength-spaced scattering elements whose orientation, geometry, or topology can be designed in a spatially varying manner to impart a desired functionality[1–5,17]. Designing metasurfaces can be challenging, however, as they often present a massive number of degrees of freedom, with millimeter to centimeter scale devices often consisting of millions to billions of individual scatterers at visible wavelengths. This poses a class of high-dimensional, fabrication-constrained optimization problems, motivating development of algorithms to efficiently traverse parameter spaces for realizing fabricable, high-performance designs.

Inverse design techniques have attracted considerable attention as a possible solution. Unlike conventional design, where intuition typically guides the process, the goal of inverse design is for a user to specify the desired performance and then use an optimization algorithm to generate a solution. Many inverse design techniques for integrated photonics[23–28] and metasurfaces[29–36] have relied on gradient descent. These gradients are typically calculated by the adjoint method[37–39], which calculates derivatives with respect to many input variables with only two simulations of a device. In photonics applications, this method has treated the electromagnetics problem in terms of an overlap between input and output modes, as this treatment yields an analytical solution to the gradient with respect to the permittivity at each point over some design volume[40,41]. While directly point-by-point optimizing the permittivity in this manner enables a large design space, converged designs often exhibit grayscale permittivity or fabrication-infeasible voids and features[42]. This necessitates spatial filters, thresholding steps, or additional merit function terms that can disrupt gradient computation to ensure realistic, fabricable designs[26,29,43,44]. Adjoint-based shape optimizations also exist, but these approaches are mostly constrained to special geometries, such as spherical or ellipsoidal Mie scatterers[45,46], by approximating transmission coefficient gradients with a polynomial proxy function[32,33] or finite differences[35], using the level set method[47,48], or expressing parameter gradients as a surface integral over the scatterer boundaries[49]. The light-matter interactions of general photonic devices, however, can be quite complicated and it is often the case that their performance metrics and design parameters are not well described by an analytical form or proxy function[50]. For example, in rigorous coupled-wave analysis[51] (RCWA), the transmission and reflection properties of a structure are functions of a global scattering matrix with elements that depend on the eigenmodes of each layer. While derivation of the adjoint fields in RCWA is possible, changing variables or scatterer geometries entails additional derivation steps for different cases, requiring extra work for a designer when switching parameterizations.

An alternative to this approach is using algorithmic differentiation[52–54], also known as automatic differentiation[55] (AD), as a means for applying the adjoint method. In AD, the output of a sequence of calculations is formulated as a computational graph in which nodes in the graph represent mathematical operations and the edges signify the flow of data from one operation to another. In this way, if each operation's derivative is known, then the derivative of the function that the graph represents can be exactly determined via the chain rule[55]. Though AD has existed for some time, its use has expanded substantially with the explosion of interest from the machine learning community with the development of artificial neural networks, giving rise to several widely used AD frameworks (e.g., TensorFlow, PyTorch, Autograd, etc.). The power of this approach is that no direct derivation of the state and adjoint equations is necessary, as this is handled implicitly in the implementation of the function of interest[52–54]. At the same time, AD-based approaches yield fast calculation of derivatives compared to both finite differences and symbolic differentiation. We emphasize that while the adjoint method is conceptually equivalent to AD in that one can explicitly derive gradients for a given scatterer parameterization, this process requires a designer to be familiar with such calculations if evaluating multiple scatterer types with varied gradient formulations. Furthermore, for many meta-optics problems, it is often unclear a priori what scatterer type is optimal, making AD a flexible technique applicable to general scatterers, enabling not only topology optimization but also general shape optimizations without requiring additional derivation steps. For example, an AD implementation of the finite-difference time-domain (FDTD) method was reported[56], enabling simulation of general photonic structures; however, the reported implementation is limited to problems with a small number of design variables in order to mitigate excessive computation times. Likewise, an AD implementation of the finite-difference frequency-domain (FDFD) method[57] has also been realized. In meta-optics design, however, RCWA is a standard method for scatterer optimization owing to its speed benefits and inherent periodic nature. For RCWA though, using AD for general scatterers or parameterized shapes would pose a complex-valued, degenerate eigenproblem[39] that is not typically differentiable[58–60] and requires perturbation theory. Though efficient AD implementations of the planewave expansion and guided-mode expansion methods were recently shown[50] with impressive results, these were restricted to handling nondegenerate eigenvalues.

In this paper, we report a generalization of existing eigendecomposition gradients to matrices with complex, degenerate eigenvalues by approximating the gradient with regularization, enabling our development of an AD implementation of RCWA using TensorFlow. We apply our developed framework to the inverse design of metagratings and metasurfaces using both topology optimization and parameterized scatterer shapes. We inverse design devices with responses multiplexed by wavelength, wavevector, and polarization. Leveraging the efficiency of backpropagation, we achieve accelerated gradient calculation[55] compared to iteration times typically encountered when using the adjoint method. Our developed method could empower designers to flexibly parameterize scatterers and realize next-generation photonic components.

**Results:**

*Generalizing Eigenproblem Gradients:*

AD has two fundamental operating modes for executing its chain rule-based gradient calculation, known as the forward and reverse modes[52,55]. To find the gradient of the output in forward mode, the derivatives of inner functions are substituted first, which consists of starting at the input nodes and moving forward towards the output while accumulating the products of each operation's derivative. Reverse mode works backwards, substituting the outer functions first, and traversing a path from the output back to the input. While both modes yield the same result, when the number of inputs is large compared to the number of outputs, reverse mode is far more efficient[52,55]. As such, we use reverse mode since we compute a single output scalar (our loss function) from at least one and potentially a multitude of input variables characterizing our unit cell(s). Leveraging AD, however, also requires all the operations in our calculation to be differentiable. Most of the calculations required for RCWA are standard operations with well-known derivatives that are readily implemented as part of an AD framework. At the heart of RCWA, however, is the solving of Fourier domain Maxwell's equations per layer, composed as the eigenequation[61]

$$\frac{d^2}{d\tilde{z}^2}\begin{bmatrix}\mathbf{s_x}\\\mathbf{s_y}\end{bmatrix} = \mathbf{\Omega}^2 \begin{bmatrix}\mathbf{s_x}\\\mathbf{s_y}\end{bmatrix}, \quad (1)$$

where $\mathbf{s}_x$ and $\mathbf{s}_y$ are the Fourier coefficients of the $x$ and $y$ components of the electric field, $\tilde{z} = k_0 z$ is the normalized $z$ coordinate along the layer thickness direction, and $\mathbf{\Omega}^2$ is a block matrix that is a function of both the wavevector expansions and the layer's permittivity and permeability. As we apply subsequent operations to the resultant eigendecomposition to ultimately calculate transmission and reflection coefficients, we need the derivatives of the eigenvectors and eigenvalues to maintain the backpropagation chain for gradients. For Hermitian matrices with nondegenerate eigenvalues, these derivatives exist[62]; however, computing eigendecomposition gradients becomes more complicated for non-Hermitian matrices as they have complex eigenvalues. Furthermore, eigenvector gradients are actually undefined in reverse mode if there are any degenerate eigenvalues[58–60]. Although workarounds do exist in forward mode[58], these would have us lose the efficiency benefits of reverse mode. These challenges preclude an exact, AD implementation of RCWA in reverse mode because while some special case scatterers present an eigenequation with distinct eigenvalues, this is not the case for general scatterer shapes and topologies.

To formulate a reverse mode RCWA implementation applicable to general scatterers, we bring together two separate innovations: generalization of the existing eigendecomposition gradient computation in TensorFlow to complex eigenvalues, and secondly a regularization scheme for approximating the reverse mode gradient for degenerate eigenvalues. The first of these innovations was previously implemented for an acoustic beamforming application[63], whereas regularization techniques for degenerate eigenvalues have been proposed separately[59,60]. To the best of our knowledge, however, these two methods have not been simultaneously applied for general eigendecomposition gradients. To aid understanding, we include here a portion of the math underpinning these two separate techniques developed in prior works[60,63]. Assume we have an eigenequation of the form

$$\mathbf{\Phi W} = \mathbf{W\Lambda}, \quad (2)$$

where the columns of $\mathbf{W}$ are the eigenvectors of $\mathbf{\Phi}$ and $\mathbf{\Lambda}$ is the diagonal matrix of eigenvalues. If we have a real output scalar function, $J = f(\mathbf{W}, \mathbf{\Lambda})$, that depends on this eigendecomposition, then when $\mathbf{\Lambda}$ is real, the reverse mode sensitivity is[62]

$$\frac{\partial J}{\partial \mathbf{\Phi}} = \mathbf{W}^{-T}\left(\frac{\partial J}{\partial \mathbf{\Lambda}} + \mathbf{F} \circ \left(\mathbf{W}^T \frac{\partial J}{\partial \mathbf{W}}\right)\right)\mathbf{W}^T, \quad (3)$$

where $\circ$ denotes the Hadamard product, $\mathbf{F}_{ij} = 1/(\lambda_i - \lambda_j)$ if $i \neq j$ and $\mathbf{F}_{ii} = 0$, and the $\lambda_i$'s are the eigenvalues. This is also the gradient provided by TensorFlow's *linalg.eigh()* function, which handles Hermitian matrices as this ensures the eigenvalues are all real. To handle complex eigenvalues, however, we need to define a complex derivative using Wirtinger calculus, which provides similar behavior to that of ordinary derivatives of real functions but for differentiable functions on a complex domain. If we assume the output scalar $J$ is real but still depends on the eigendecomposition when we have complex $\mathbf{\Lambda}$, we can instead differentiate with respect to $\mathbf{\Phi}^*$ using the Wirtinger derivative. This yields

$$\frac{\partial J}{\partial \mathbf{\Phi}^*} = \mathbf{W}^{-H}\left(\frac{\partial J}{\partial \mathbf{\Lambda}^*} + \mathbf{F}^* \circ \left(\mathbf{W}^H \frac{\partial J}{\partial \mathbf{W}^*}\right)\right)\mathbf{W}^H, \quad (4)$$

as derived previously[63]. In both the real and complex versions, if there are any degenerate eigenvalues, then both $\mathbf{F}$ and the gradient are undefined. To circumvent this, we utilize a regularization scheme leveraged previously[60], applying a Lorentzian broadening where we instead set

$$\mathbf{F}_{ij} = \frac{\lambda_i - \lambda_j}{(\lambda_i - \lambda_j)^2 + \varepsilon}, \quad (5)$$

with $\varepsilon$ being a small real number, a hyperparameter that produces a small error in $\mathbf{F}_{ij}$ but regularizes our gradient calculation for complex, degenerate eigenvalues. We select $\varepsilon$ by adjusting its value to minimize the gradient error compared to a finite difference calculation (see Supplementary Note 1 and Figure S1).

*Inverse Design Framework:*

With a reverse mode implementation of general eigendecomposition, we can now construct an end-to-end differentiable RCWA inverse design framework using AD. Our approach is shown schematically in Figure 1. The key aspects of the framework are the hyperparameters and optimization variables, the RCWA implementation itself, the input excitations organized as a batch tensor, and the loss function (i.e., the figure of merit) to optimize. The hyperparameters define the system under consideration and form the base tensor that represents the shape of our network through which data will flow (e.g., how many real space points are used to represent the permittivity, how many dielectric layers our structure has, etc.). The optimization variables depend on the unit cell parameterization or topology. While our framework enables the permittivity to vary freely within each unit cell, we can just as easily define a shape optimization with respect to the geometric parameters of a scatter. This includes optimizing the diameter and thickness of a conventional, nanopost-based grating or metasurface without having to derive any complicated state and adjoint equations that depend upon the variable selection. With a given parameterization, we can then calculate the permittivity of our unit cell and feed that unit cell into the RCWA model. The RCWA implementation comprises the core of the framework, modelling the physics of the unit cell in a forward pass. This includes generating convolution matrices, solving the Fourier domain eigenequation for the modes in each layer of the structure, finding the global scattering matrix via a Redheffer star product of the individual layer scattering matrices, and extracting the transmission and reflection coefficients for the unit cell(s). This yields a black box defining an input-output relationship for a given structure. Then, for a batch of input excitations (e.g., for varying wavelength, polarization, or wavevector) that feed into our tensor network, we determine the state of the output light when it is incident on the unit cell(s). Finally, this output is passed into a scalar loss function, which could simply be the reflectance or transmittance, or more complicated metrics based on responses multiplexed by the input as desired by the user. Alternatively, our tensor network could constitute an array of many unit cells as part of a full metasurface (Figure 1), and the loss function could be based on properties of the diffracted wavefront, such as maximizing the field magnitude at a single point for a lens.

*Benchmarking:*

To benchmark our framework, we first evaluate the accuracy of our gradient calculation as this is critical in its ability to perform optimizations. For this purpose, we consider a periodic grating of 633 nm tall $TiO_2$ cylinders on an $SiO_2$ substrate illuminated with 633 nm light as a model system and compute the partial derivative of the reflectance with respect to the duty cycle (i.e., diameter normalized by a grating pitch of 443 nm) using our AD technique and compare this to that found by the finite difference method (Figure 2a). The gradient computed by our technique agrees well with that from the finite difference approximation, even on resonance (Figure 2a inset). The mean fractional error for the computed gradient in this model system was 0.56%, demonstrating how closely our method agrees with the true gradient and validating the accuracy of our approach (see Supplementary Note 1 and Figure S1). The validity of the underlying RCWA solver itself was also evaluated, which was found to be in strong agreement with results produced by the commonly used S4 package[64] (see Supplementary Note 2 and 3 and Figures S2 and S3).

To assess the computational cost, we consider the time per optimization iteration, as this is the most significant contributor to the total time for adjoint methods. For the adjoint method, each iteration usually costs twice the forward simulation time, as both forward and adjoint simulations are conducted to determine the gradient. In the worst case, the discrete adjoint method would cost six times the forward simulation time per iteration, an upper bound provided by a theoretical result[52,54,65] that backpropagation cannot exceed five times the cost of a forward run. In practice, however, AD methods often determine gradients at a small multiple of the cost of forward simulation[55]. Our framework performed far better than this worst case (Figure 2b), and as the number of optimization variables increases with the number of metasurface scatterers, we achieved average iteration times as low as 1.24 times the forward simulation time, 38% less than the typical time of two forward simulations for adjoint methods. This reduction in relative iteration time compared to the adjoint method is not possible for all functions but is achievable in our case by reusing the pre-computed eigenvectors and eigenvalues from the forward pass. As the most time-consuming RCWA computation is the eigendecomposition step and as the corresponding gradients depend on these values, storing the results as is required for reverse mode, provides a significant speedup. The absolute time per iteration depends on the designer's hardware but as our method is implemented in TensorFlow, a standard deep learning platform, the computation is highly amenable to acceleration using graphics processing units (GPUs), tensor processing units (TPUs), or other dedicated hardware. Figure 2c shows the speedup factor when the same optimization is performed using a GPU compared with a CPU, demonstrating 14x and 24x speedups for the tested shape and topology optimizations, respectively.

*Single-layer Metagratings:*

One of the key benefits of our approach is the flexibility it offers in terms of unit cell structure. To demonstrate this, we applied our method to the inverse design of high-performance, single layer metagratings based on various unit cell types. We first consider a frequently encountered unit cell comprising a cylinder of $TiO_2$ on an $SiO_2$ substrate (Figure 3a), the dimensions of which we want to adjust to optimize the reflectance at discrete wavelengths. With only the cylinder diameter as a free variable, at an excitation wavelength of 633 nm, we maximize the reflectivity to more than 99.8% compared to an initial value of 62.7%. The learning curve in Figure 3a exhibits minor, decaying oscillations as the reflectivity approaches convergence, arising from successive hopping about the locally optimal diameter. Physically, the change from the initial to final grating design duty cycle manifests as a blueshift (Figure 3b). Without explicitly deriving any gradients, we can change variables to support alternative unit cell parameterizations, enabling inverse design of a high extinction ratio (22 dB) polarizer based on rectangular scatterers (Figure 3c-d), as well as a topology-optimized, silicon metagrating that deflects 1550 nm light to 75° with absolute and relative efficiencies of 77% and 95% respectively (Figure 3e-f).

*Multilayer Devices:*

While flat optics based on single layer metagratings have a broad range of applications, they have a limited number of degrees of freedom. Multilayer devices can increase the degrees of freedom and exploit mutual interactions and coupling amongst modes in stacked layers, potentially enhancing the desired performance. As RCWA is intended for such design problems, our approach lends itself well to multilayer devices, allowing a designer to select at will the number of unit cell layers in addition to their topology or scatterer shape. We first demonstrate this for shape-optimized, bilayer $TiO_2$ cylinders (Figure 4a) whose reflected power is multiplexed by the light's incident angle. By optimizing with respect to the cylinder thicknesses and diameters in the stacked layers, the converged design achieved high-efficiency reflection of 633 nm light at 2.5° while letting it pass through unattenuated at 7.5° (Figure 4b).

The same multilayer capability is also extendable to stacked, topology-optimized metagratings. Here, we consider layers of $TiO_2$ with interstitial 1.5 refractive index material (e.g., filled with $SiO_2$ or a polymer). By optimizing each layer's transverse permittivity profile, we inverse design a 10-layer, volumetric grating that simultaneously deflects red (700 nm), green (520 nm), and blue (420 nm) light to 60° with high efficiency. We achieve this by performing a maximin optimization over the three wavelengths to uniformly boost their diffraction efficiencies and by selecting a 2.425 μm pitch that enables us to select separate diffraction orders for each wavelength that coincide with the same deflection angle. With each wavelength achieving a minimum of 45% absolute efficiency and relative efficiencies all more than 67%, this approach could find applications in augmented reality waveguides, where RGB light must be simultaneously deflected at large angles exceeding the critical angle.

*Full Metasurfaces:*

In addition to optimizing the unit cell of periodic gratings, our method supports full metasurface optimizations wherein arrays of scatterers may impart spatially varying transformations on wavefronts. The applicability of our method to these devices is based on the local phase approximation, in which the slow variation of scatterers with position allows us to approximate each scatterer with the response of one positioned in a periodic lattice. This enables us to capture the essence of each scatterer with the RCWA-computed transmission and reflection coefficients. By coupling our differentiable RCWA module to an AD-based implementation of the band-limited angular spectrum method[66], we form a full pipeline in which we can optimize electric fields diffracted by a spatially varying metasurface with respect to an arbitrary parameterization for our unit cells. Here, we define a shape optimization based on a unit cell comprising 4 coupled elliptical $TiO_2$ nanoposts on an $SiO_2$ substrate (Figure 5a), with each nanopost characterized by their respective major and minor axes. This unit cell selection offers multiple degrees of freedom (8 per unit cell) that can mitigate possible convergence to poor local optima and enables asymmetric scatterer designs that can support polarization discrimination.

We first demonstrate the inverse design of monochromatic lenses at 633 nm for normal incidence. We maximize the electric field magnitude at a single point on the optical axis corresponding to the desired focal length for designs over a range of different f-number values. Figures 5b and 5c show the learning curve and optimized intensity profile at focus for a designed f/1 lens, while Figure 5d shows intensity profiles along the optical axis for several different f-number designs, demonstrating convergence over a wide range of focal lengths. As our optimization is local and we must contend with the accuracy limitations of the local phase approximation, the intensity profiles of Figure 5d computed using FDTD indicate that even for an optimized design, a nonnegligible portion of light can still be directed to locations away from focus, sometimes including secondary and tertiary focal spots[67,68]. While this limits the mean focusing efficiency of the optimized devices to 16% (Figure 5e), we still achieve focal spot sizes that are less than that predicted by the Abbe limit (Figure 5f); however, we emphasize that this is not beating the diffraction limit by any reasonable metric[69], but rather this can be understood as the focused beam profile deviating from that of the Airy disk and additional power being shifted into outer intensity rings, increasing the peak sidelobe ratio of the focal profile (Figure 5g).

To demonstrate the polarization capability of our framework for an aperiodic device, we then inverse designed a polarization-multiplexed metalens that directs vertically and horizontally polarized incident light to separate focal spots (Figure 6a). We leveraged the same coupled elliptical nanopost unit cell parameterization (Figure 5a) as for the monochromatic lenses, but for this design we consider two incident linear polarization states as a batch tensor. Our loss function is based on the product of electric field magnitudes at the desired focal spots (Figure 6b) for each respective polarization state, which are located symmetrically with respect to the optical axis in the y-z plane. FDTD simulations of the optimized structure validate the polarization-controlled focusing (Figure 6c-e).

**Discussion:**

Our framework applies to a broad range of nanophotonic structures and is especially well suited to the optimization of periodic, layered structures owing to the periodic nature of RCWA. It is also applicable to the design of metasurfaces as long as the local phase approximation is valid, which is commonly assumed in the metasurface community but works best for high index contrast structures in which coupling among adjacent elements is weak[8] and is limited for low index designs[70,71]. The approach is also extendable to a broader range of structures by exploiting overlapping domains of unit cells[72].

Unlike the adjoint method, our AD-based approach lends itself well to arbitrary unit cell types. This enables not only topology optimization, where like existing adjoint methods we can enforce fabrication constraints by means of feature blurring and thresholding, but also shape optimizations where fabrication feasibility is inherently satisfied by limiting the ranges of a given shape's geometric parameters. An additional key benefit is that in using AD, fabrication constraints are applied in a differentiable manner by directly incorporating their effect as an operator applied to the permittivity function within the backpropagation chain. This avoids the need to apply them intermittently outside of the variable update steps. Furthermore, multilayer devices can be easily accommodated using our method. By supporting arbitrary unit cells, we empower the designer to select, and later modify if desired, whichever parameterization or topology they deem best fit to their application without having to work through any explicit gradient derivation for different types of scatterers.

Though our eigenvalue regularization yields an approximate gradient, we find that in practice the technique works well, as demonstrated by its strong agreement with results from the finite difference method (Figure 2a) and the convergence results of the optimizations in this paper. The speedup offered by our implementation in terms of iteration time relative to a forward run (Figure 2b) could also significantly reduce design time. Though the absolute iteration time depends on the available hardware, our TensorFlow implementation enabled large speedups when switching from a CPU to a GPU (Figure 2c). While in this work we use the Adam algorithm[73] to update variables given the calculated gradients, an additional key benefit of our platform is that it is readily adaptable to support alternative, gradient-based optimization algorithms or to incorporate more sophisticated fabrication robustness routines. Our RCWA implementation could plug into other meta-optical optimization algorithms in place of alternative RCWA simulators that are not already algorithmically differentiable. This could reduce iteration time and increase design flexibility while leveraging already existing and mature, application-specific optimization routines.

In this paper, we described the development of a photonic inverse design framework based on an AD implementation of rigorous coupled-wave analysis. We generalized TensorFlow's existing eigendecomposition gradients to complex, degenerate eigenvalue problems by synthesizing advances in this domain from two separate works[60,63]. This yielded an approximate gradient that supported a fully differentiable, end-to-end implementation of RCWA for a variety of photonic design problems. This enabled arbitrary shape parameterization and topology optimization of unit cells and we demonstrated applicability of the method to engineering the response of metagratings as a function of wavelength, angle, and polarization. We then coupled our RCWA implementation to a subsequent Fourier optics AD diffraction module, enabling optimization of full metasurfaces within the local phase approximation. Our approach could find widespread use in designing periodic structures and high index metasurfaces, with the potential for supporting a broader range of photonics design problems with additional modifications to our framework. With the growing demand for high-performance photonics, we believe the flexibility of our approach and its relative speed in cost per iteration will present a competitive inverse design method for photonic devices. We note that during peer review, a related work[74] was published that describes a method for exactly differentiating the scattering matrices encountered in RCWA via an elegant circumvention of the eigendecomposition differentiation step. Our approach, however, is still unique in its leveraging of automatic differentiation and regularization, offering a simple and open-source framework for inverse design of a diverse range of meta-optics.

**Methods:**

*Optimization Settings:* All optimizations were conducted using the TensorFlow implementation of RCWA, which for the case of full metasurfaces is followed by computation using a TensorFlow version of the band-limited angular spectrum method[66]. The implementation is based on a chain of differentiable tensor operations acting on a base input tensor with six dimensions, $(batch, pixels_x, pixels_y, N_{layer}, N_x, N_y)$. $N_x$ and $N_y$ represent the points in the real space cartesian grid that constitute each unit cell, $N_{layer}$ corresponds to the different stacked layers in the structure, $pixels_x$ and $pixels_y$ access the different scatterer positions of a full metasurface and are both set to 1 for a periodic simulation, and over the $batch$ dimension the input conditions vary (e.g., polarization, wavelength, and wavevector). The cartesian real space grid on which the permittivity and permeability for each unit cell are sampled is converted into a convolution matrix, the size of which depends on the number of Fourier harmonics used, which is a hyperparameter for the optimization.

2D grating simulations in this work utilized 121 Fourier harmonics, 11 along each axis, with the permittivity profile discretized on a $512 \times 512$ point cartesian grid. For 1D grating simulations (Figure 4e-f) we utilize 81 harmonics in the variable direction and 1 harmonic along the axis where the permittivity profile is unchanging, and we discretize into a $512 \times 1$ point cartesian grid. In Figure 3a-b, the cylinder is initialized with a duty cycle of 60%, thickness of 632 nm, and pitch of 443 nm. The polarizing grating of Figure 3c-d was initialized with a duty cycle of 40%, pitch of 474 nm, and thickness of 632 nm. The topology optimization of Figure 3e-f begins with a random permittivity profile sampled in range from 1.0 to 6.76 (the value for $TiO_2$). The bilayer grating of Figure 4a-b starts with top and bottom thickness of 632 nm, and top and bottom duty cycles of 70%. The multilayer topology optimization (Figure 4c-d) has all layers starting as random permittivity distributions in the 2.25 to 6.76 range (the values for $SiO_2$ and $TiO_2$).

Full metasurface optimizations used 25 Fourier harmonics per scatterer unit cell, 5 along each axis, with each scatterer's permittivity sampled on a $256 \times 256$ point grid. Analysis of the diffraction efficiency convergence as a function of number of harmonics can be found in Supplementary Note 4 and Figures S4 and S5. Each metasurface consisted of $31 \times 31$ scatterers with a fixed pitch of $0.7\lambda$, yielding apertures with a width of 13.7 μm. In Figures 5 and 6, all unit cells are uniformly initialized to a duty cycle of 35% for all axes and the thickness and pitch are fixed at 632 nm and 443 nm, respectively. To calculate the efficiencies in Figure 5, we take the ratio of the power at the focal spot within the area encompassed by a circle with a radius of 3 times the full width at half maximum to the power incident on the metasurface.

For shape optimizations, the permittivity in a layer is computed from the parameters describing the cross-sectional geometry of a layer (e.g., scatterer diameter or width as opposed to the thickness). To make this differentiable, we first evaluate the polynomial describing the shape boundary zeros of our scatter at all grid positions in the layer (e.g., $x^2 + y^2 - r^2$ for a circle). This is then multiplied by a large number, a hyperparameter, and is passed as an argument to a sigmoid function that acts as a differentiable threshold. For negative values, inside the boundary, this gives 0 while for positive values, outside the shape, this gives 1 so that we then have only two permittivity values (the scatterer and background permittivity values). As the sigmoid is continuous, for positions along the shape boundary, there can be intermediate permittivity, but in practice because of the large sigmoid argument scaling hyperparameter, these intermediate values would be confined to just the shape boundary and have a negligible effect on the performance of our designs. The focal distance for metasurfaces as well as material refractive indices for shape optimizations, substrate thickness, unit cell pitch, and the Lorentzian broadening parameter for eigendecomposition gradients are also hyperparameters of the tensor network. All optimizations used the Adam algorithm with varied learning rates, number of iterations, and variable initializations. Example codes are provided on the GitHub repository.

For topology optimization of single 2D unit cells (Figure 3e-f), we again break the design into a $512 \times 512$ point cartesian grid. At each iteration, we apply a differentiable blurring operation by convolving the density (i.e., the permittivity normalized to be within the 0 to 1 range) with a blur kernel[29] with a radius related to the minimum fabricable feature size. The structure is then passed through a differentiable thresholding step to gradually push it towards a binary structure at convergence. At the end of the optimization, the final structure is passed through a final binary threshold to eliminate any remaining grayscale features. The blurring and thresholding operations at each iteration are implemented as differentiable operations within the computational graph, rather than as operators applied after gradient calculation and variable updates. This enables us to directly backpropagate through these fabrication constraint-enforcing operations. For 1D topology optimizations (Figure 4e-f), the same blurring and thresholding operations are applied as in the 2D case.

*Physical Validation:* The TensorFlow-implemented rigorous coupled-wave analysis framework was validated against results produced by S4[64], showing strong agreement (see Supplementary Note 2 and 3 and Figures S3 and S4). For full metasurfaces, the optimized structures were simulated in Lumerical FDTD and the resulting transmitted near-field data was propagated to subsequent planes using the band-limited angular spectrum method[66], yielding the intensity profiles present in Figures 5 and 6.

*Computation:* Full metasurface simulations conducted here were run on Amazon EC2 (Deep Learning AMI, Ubuntu 18.04, Version 27.0, TensorFlow version 2.1.0, instance types g4dn.xlarge and g4dn.8xlarge) while grating optimizations were performed on the Python 3 Google Compute Engine Backend available in the Google Colab environment, where we used a Tesla V100 GPU and an Intel Xeon CPU with 26 GB RAM. The shape optimizations in Figure 2 used arrays of 633 nm tall $TiO_2$ cylinders on an $SiO_2$ substrate with a pitch of 443 nm illuminated with 633 nm light and $5 \times 5$ harmonics. For the topology optimizations of Figure 2B-C, we used $5 \times 5$ harmonics and let the permittivity vary between that of $TiO_2$ and $SiO_2$. For both the shape and topology optimization benchmarking evaluations, we utilized a $128 \times 128$ cartesian grid per unit cell.

**Acknowledgements**: This research was supported by NSF-1825308, DARPA (Contract no. 140D0420C0060), the UW Reality Lab, Facebook, Google, Futurewei, and Amazon. A.M. is also supported by a Washington Research Foundation distinguished investigator award. We also acknowledge useful input from Dr. Alan Zhan.

**Author Contributions:**

S.C. formulated the approximate reverse mode sensitivity for eigendecomposition of general matrices, developed the TensorFlow simulation and optimization framework, and compiled the optimization results. A.M. supervised the project and helped S.C. write the paper.

**Competing Interests:**

All authors declare no competing interests.

**Data Availability:**

The data that support the findings of this study are available from the corresponding authors upon reasonable request.

**Code Availability:**

Code available at https://github.com/scolburn54/rcwa_tf

**Figure Legends:**

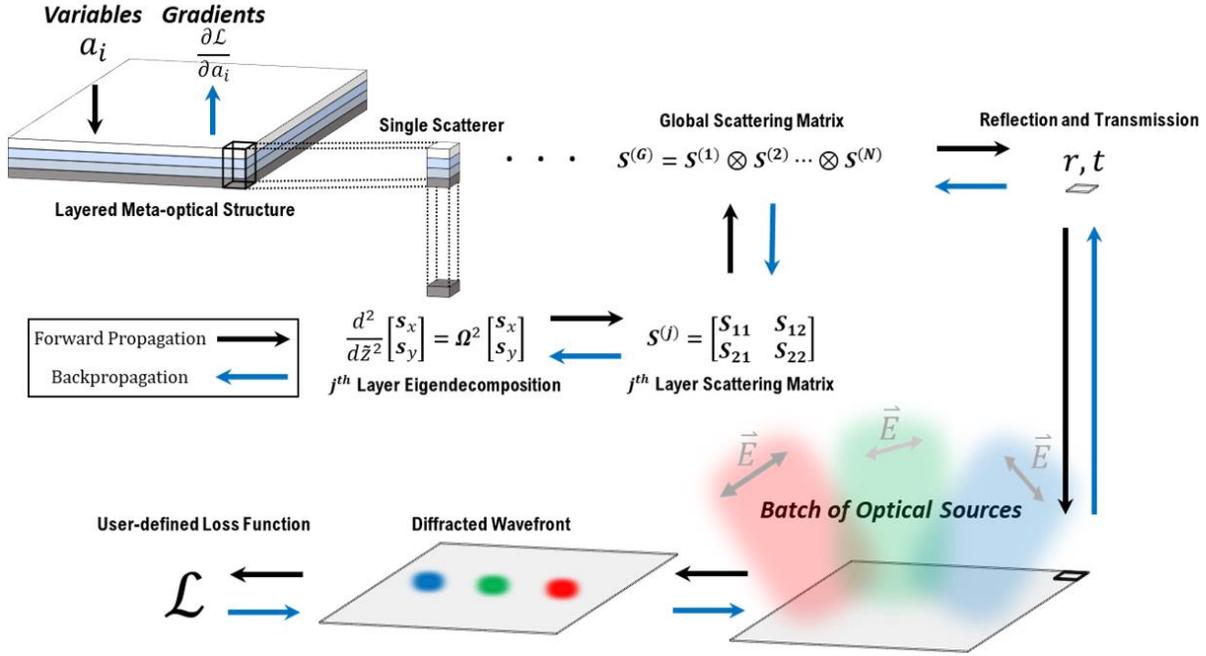

**Figure 1.** Meta-optical Computational Graph Schematic. Our framework treats a photonics problem like a neural network, where structural variables $a_i$ are trainable weights that determine the structure of a layered meta-optical element, inputs ($\vec{\mathbf{E}}$) are a batch of optical sources, and the user-defined loss function ($\mathcal{L}$) captures the desired performance of the device. A sequence of differentiable tensor operations acting on the structural variables $a_i$ forward propagate (black arrows) data through the network, solving the eigenproblem posed by the matrix wave equation in which $\mathbf{\Omega}^2$ is a block matrix depending on the layer properties and wavevector expansions, $\tilde{z}$ is the normalized coordinate along the layer thickness direction, and $s_x$ and $s_y$ are the Fourier domain electric field coefficients for x and y. This is performed for each layer, the number $N$ of which is determined by the user and dictates the range of $j$, which we use here to denote the $j^{th}$ layer. The solved eigenequation for each layer enables us to calculate the scattering matrix $\mathbf{S}^{(j)}$ for the $j^{th}$ layer, where the $\mathbf{S}_{11}$, $\mathbf{S}_{12}$, $\mathbf{S}_{21}$, and $\mathbf{S}_{22}$ elements denote the standard scattering matrix components for characterizing ingoing and outgoing fields. The results of these separate layer scattering matrices, going up through the $N^{th}$ layer, are all combined via a Redheffer star product to form $\mathbf{S}^{(G)}$, where the $G$ superscript denotes that this is the global scattering matrix of the full system. From $\mathbf{S}^{(G)}$, the reflection ($r$) and transmission ($t$) coefficients are determined, which then constitute the electric fields to be diffracted in combination with the inputs $\vec{\mathbf{E}}$. Acting upon this batch of input optical sources with varying wavelength, wavevector, and polarization, the data is finally passed to the scalar, user-defined loss function ($\mathcal{L}$) that is to be minimized. Data is then backpropagated (blue arrows) to enable computation of gradients $\frac{\partial \mathcal{L}}{\partial a_i}$ to update the meta-optical structure.

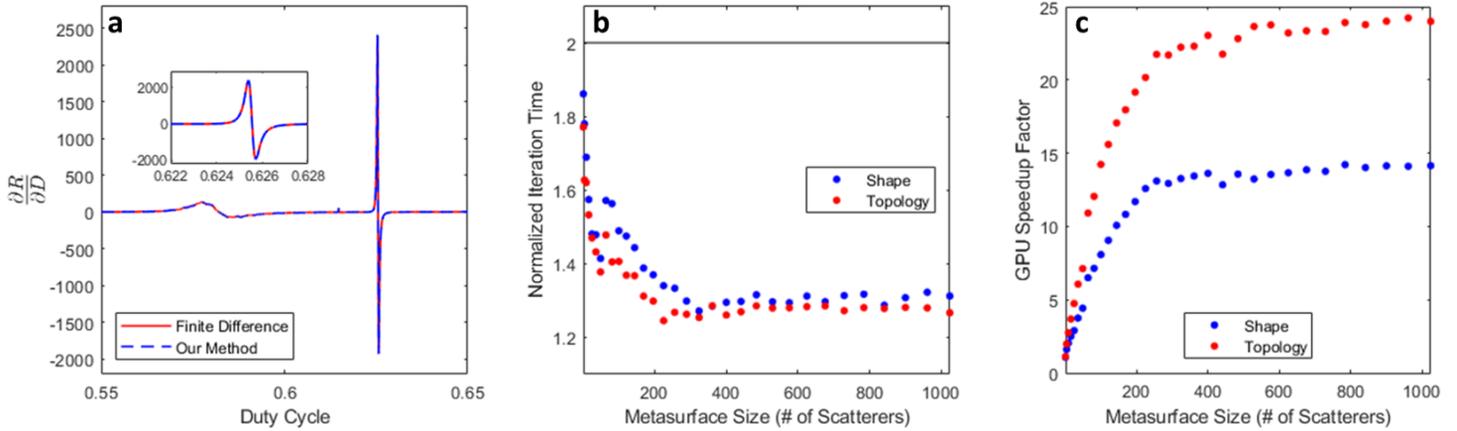

**Figure 2.** Evaluating Gradients and Iteration Time. (a) Our method of computing the partial derivative of the reflectance $R$ with respect to the duty cycle $D$ (i.e., the diameter to pitch ratio) of a cylindrical nanopost TiO$_2$ grating agrees well with the results of the finite difference method, even on resonance (zoomed in the inset). (b) For both topology and shape optimizations, our method represents a marked speedup in iteration time relative to the time for a respective forward run owing to our reuse of the costly eigendecomposition step from the forward pass during gradient calculation. In both cases, our method achieves normalized iteration times less than those typical of the adjoint method, denoted by the black horizontal line at twice the forward simulation time. (c) Our framework is amenable to GPU acceleration, enabling significant reductions in computation time compared to running the same optimization on a CPU. As the number of scatterers simulated increases, the speedup factor eventually saturates, with topology optimizations typically obtaining a greater benefit from a GPU. In both (a) and (b), the points shown are mean values, the errors for which are omitted from the plot as they are smaller than the symbol size.

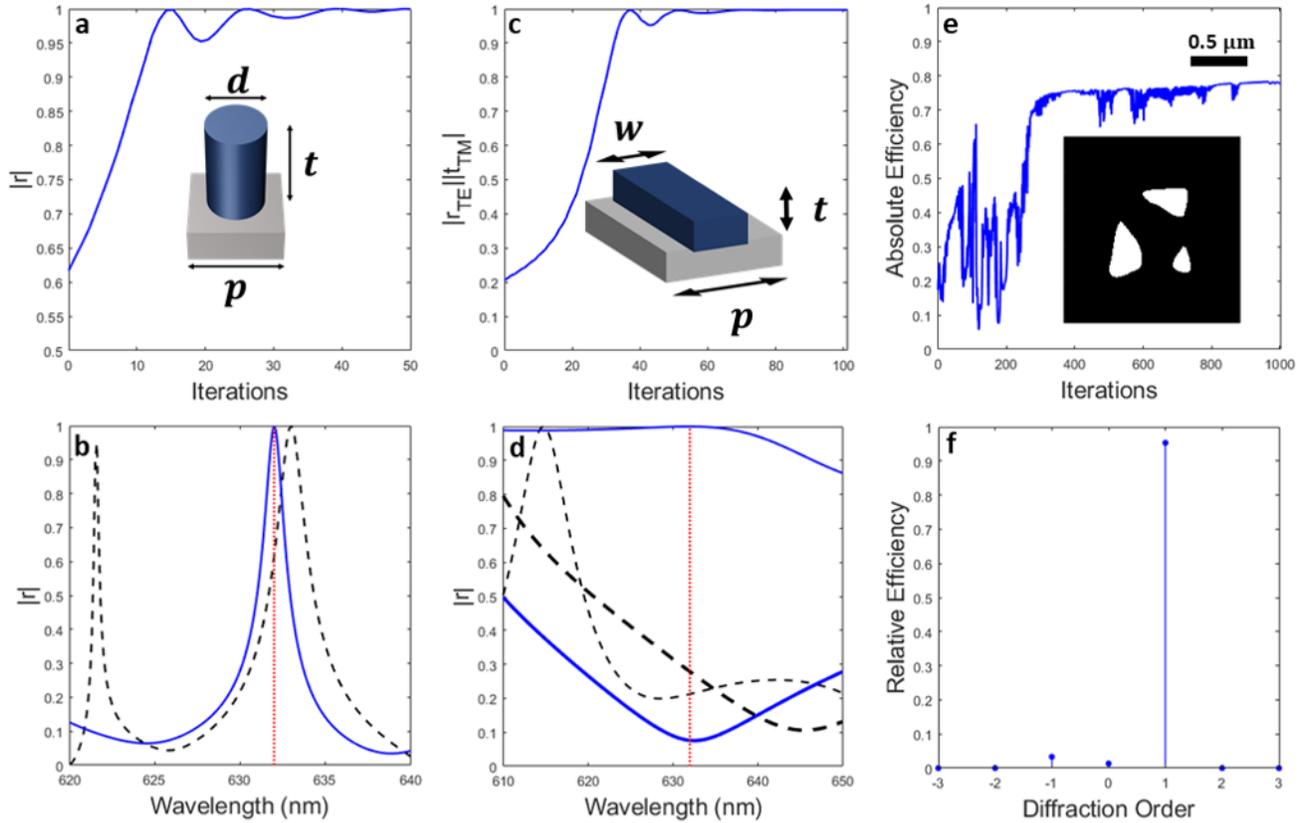

**Figure 3.** Optimization Results for Single Layer Metagrating Designs. We maximize the reflectivity (a) of a TiO$_2$ cylindrical nanopost (a inset) by optimizing the diameter, inducing a blueshift of the resonance (b), where the black dashed and blue solid curves are the initial and optimized designs respectively and the red dotted line denotes the design wavelength of 632 nm. In (a), $d$, $t$, and $p$ denote the diameter, thickness, and pitch of the grating as shown. In (c), we maximize the transverse electric (TE) reflectivity $r_{TE}$ while maximizing the transverse magnetic (TM) transmission $t_{TM}$ of a TiO$_2$ rectangular line-based unit cell (c inset) at a design wavelength of 632 nm. In (d), the results of this optimization are shown where the red dotted line denotes the design wavelength, the black thin (thick) dashed line represents the initial TE (TM) reflectivity while the blue thin (thick) solid line stands for the optimized TE (TM) reflectivity. The optimized design achieved an extinction ratio of transmitted TE to TM light of 22 dB. In (a), $w$, $t$, and $p$ denote the width, thickness, and pitch of the grating as shown. The dashed and solid lines in (d) denote TM and TE polarizations respectively, while black and blue indicate the initial and optimized designs respectively. We also topology optimize (e) a single-layer silicon grating at 1550 nm (e inset shows the permittivity profile, with silicon in white and SiO$_2$ in black), achieving an absolute efficiency of 77% of light directed into the first diffraction order (75° deflection) and a relative efficiency of 95% (f).

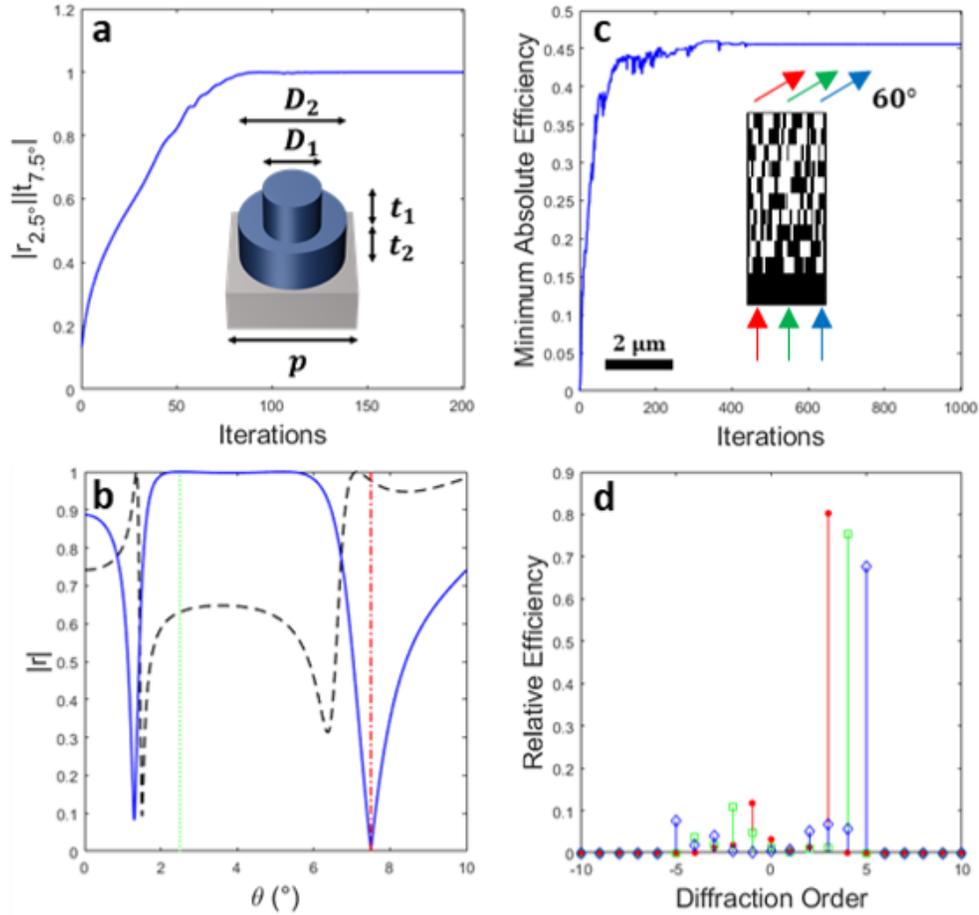

**Figure 4.** Optimization Results for Multilayer Metagrating Designs. We maximize the reflectivity at an input angle $\theta$ of 2.5°, $r_{2.5°}$, and the transmissivity at 7.5°, $t_{7.5°}$, (a) of a bilayer TiO$_2$ cylinder (a inset) by simultaneously adjusting the bottom and top layer thicknesses and diameters. In the (a) inset, $D_1$ ($D_2$) and $t_1$ ($t_2$) denote the top (bottom) cylinder diameter and thickness respectively, whereas $p$ is the pitch of the grating. In (b), the optimized reflectivity (blue solid line) shows the resulting high reflectivity at an input angle $\theta$ of 2.5° (green dotted line), whereas it exhibits a prominent dip at 7.5° (red dot dash line) corresponding to the intended high transmissivity at that angle, whereas the black dashed line (initial design) performs poorly at both angles. In (c), we simultaneously optimize the diffraction of red, green, and blue wavelengths to 60° by maximizing the minimum diffraction efficiency of the three wavelengths based on multilayer TiO$_2$ grating (c inset with white as TiO$_2$ and black as 1.5 index material). To achieve diffraction at the same angle, we fix the pitch at 2.425 μm and optimize the direction of red, green, and blue light into the 3rd, 4th, and 5th diffraction orders respectively, reaching relative efficiencies exceeding 67%. In (d), the relative efficiencies for red (dots), green (squares), and blue (diamonds) are plotted as a function of diffraction order.

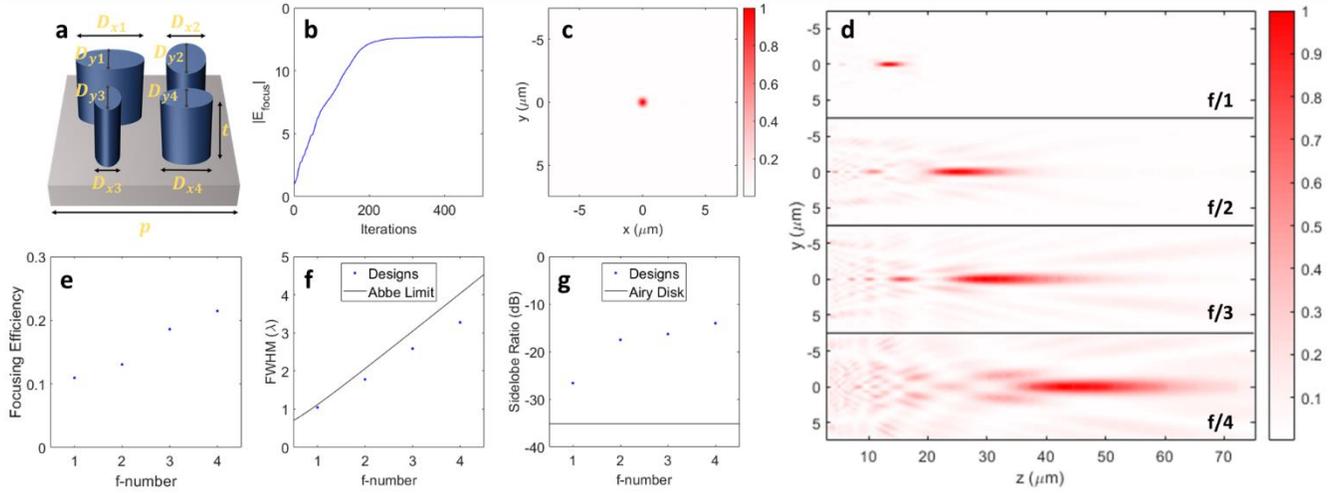

**Figure 5.** Optimization Results for Metalenses. (a) The unit cell is based on four coupled TiO$_2$ elliptical nanopost resonators, each defined by their respective major and minor axes. The labels $D_{xi}$ and $D_{yi}$ denote the $x$ and $y$ semi-diameters of the $i^{th}$ nanopost of the 4 resonators in the unit cell respectively. The labels $t$ and $p$ denote the thickness and pitch of the unit cell. An example learning curve where we optimize the electric field magnitude at the desired focal spot, $|E_{focus}|$, is shown in (b) and its corresponding normalized intensity profile computed via finite-difference time-domain (FDTD) simulation at the focal plane is presented in (c). Normalized intensity cross sections along the optical axis for several different optimized metalenses with different f-numbers, with the near-field data computed via FDTD, are shown in (d). Focusing efficiency (e), FWHM at best focus (f), and the peak sidelobe ratio (g) are calculated from the FDTD-simulated optimized designs of (d).

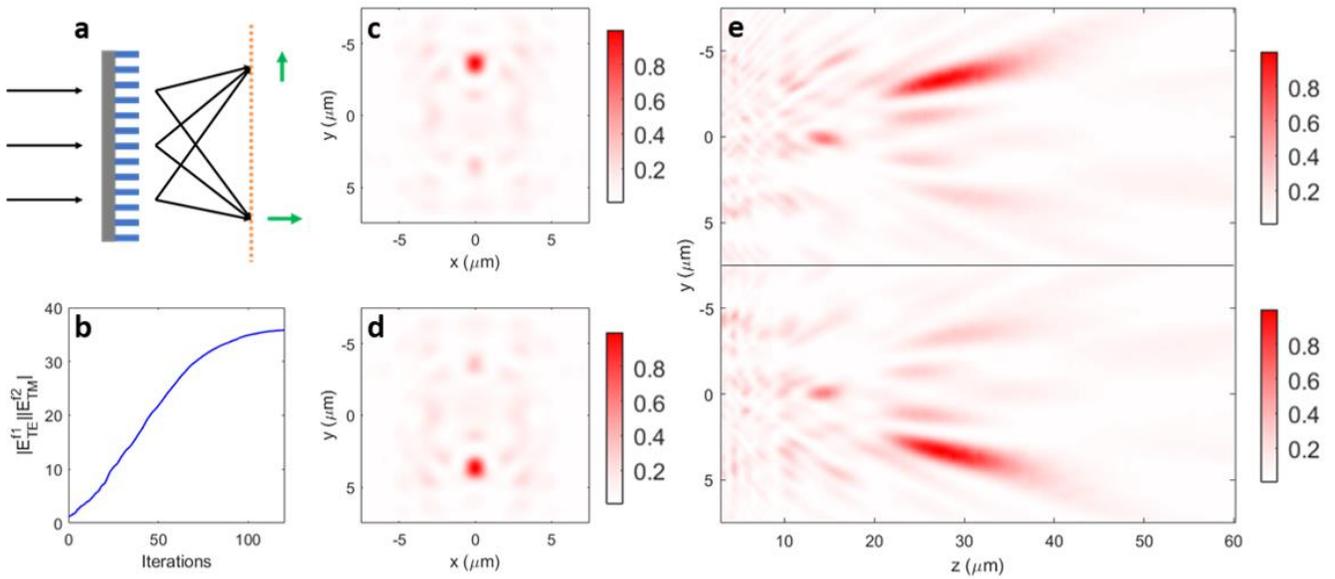

**Figure 6.** Optimization Results for Polarization-multiplexed Metalenses. (A) Schematic of a cross section of the proposed device where incident light is focused to two different focal spots controlled by the linear polarization state of the input. The black arrows indicate the direction of light propagation, the orange dotted line designates the focal plane, and the green arrows represent the orientation of the distinct horizontal and vertical polarization states of the foci. The product of the electric field magnitudes at the two different focal spots is optimized (B) and the resultant intensity at the focal plane for vertical (C) and horizontal (D) polarizations exhibit distinct focal spots. Focusing in the y-z plane for each polarization state for the optimized design is shown in (E), with near-field data computed using finite-difference time-domain simulation. TE and TM in (B) stand for the transverse electric and transverse magnetic polarization states, whereas $E^{f_1}$ and $E^{f_2}$ denote the electric fields at the two optimized focal spot positions.